\documentclass[aps,prl,twocolumn,groupedaddress]{revtex4-1}
\usepackage[dvipdfmx]{graphicx}
\usepackage{subfigure}
\usepackage{here}
\usepackage{longtable}
\usepackage{amsmath}
\usepackage{comment}

\begin{document}

\title{First-Principles Study on Preferential Energetics of Mg-based Ternary Alloys Revisited by Short-Range Order in Disordered Phases}

\author{Ryohei Tanaka and Koretaka Yuge}
\affiliation{Department of Materials Science and Engineering, Kyoto University, Kyoto, 606-8501, Japan}

\date{\today}

\begin{abstract}

To investigate the formation of Mg-based long-period stacking ordered (LPSO) structure, we systematically study 
the preference of the short-range order (SRO) in metastable disordered phases of Mg-RE-Zn (RE = Y, La, Er, Ho, Dy, Tb) and Mg-Gd-Al ternary alloy systems 
through first-principles calculation. RE-Zn (RE = Y, Er, Ho, Dy, Tb) and Gd-Al pair clusters' probability showed the tendency of increasing. In contrast, La-Zn pair clusters' probability is not increasing,
 whose system dose not form LPSO structure. This preference of SRO in disordered phases of Mg-based ternary alloys certainly indicates that peculiar L1$_2$-type ordering in LPSO as well as possibility of LPSO formation should have strong correlation with SRO tendency of energetically competitive disordered phases.
 
 \end{abstract}

\pacs{}

\maketitle
\section{1. Introduction}

Mg-based long-period stacking ordered (LPSO) structures are considered as next-generation lightweight structural alloys for their outstanding tensile strength and ductility~\cite{Kawamura}. 
These novel properties of Mg-based LPSO structures are considered to be closely related with the formation of the periodical stacking faults and clusters consisted of dilute substitutional elements.
Thus,  in order to clarify the relationship between their formation and their physical properties, a number of experimental and theoretical studies have been carried out.  In experimental studies, using in-situ synchrotron radiation small angle X-ray scattering, Okuda $et$ $al$. clarified a hierarchical phase transformation of Mg-Y-Zn LPSO structure; the clusters distributed randomly and finally transformed into a microstructure with particular distances, which introduces the periodical stacking faults into Mg-based hcp stacking sequence that required for the 18R LPSO structures~\cite{Okuda}.  In theoretical approach, on the other hand, it is still difficult to efficiently predict structures and physical properties of multicomponent alloy because we need to take manybody interaction into consideration, resulting in a large computational cost. Therefore, various approaches are proposed. 
Kimizuka $et$ $al$. constructed an on-lattice effective multibody potential model and successfully predicted the clustering of solute atoms and in-plane ordering of L1$_{2}$ clusters in dilute Mg-based alloy system using Monte Carlo simulation~\cite{Kimizuka1}~\cite{Kimizuka2}. They predicted that the repulsive interaction among solute clusters is a contributing facto for a the two-dimensional medium-range ordering of clusters.

In our previous study~\cite{Tanaka}, focusing on the stability of disordered phases competing with ordered phases, the preferential energetics of Mg-Y-Zn ternary alloy was investigated based on special quasirandom structure (SQS)~\cite{Zunger}. We clarified the thermodynamic stability of disordered phases of Mg-Y-Zn alloys, especially the formation free energy of SQSs, the effect of phonon to the stability in terms of bulk modulus and the preference of the interfacial energy when introducing stacking faults into hcp phases, which should be a fundamental prerequisite for the acceleration of forming characteristic LPSO structure. 

These obtained values by using SQSs correspond to physical quantities in disordered phases at high temperature limit. In order to clarify the formation of Mg-based LPSO structure, it is therefore essential to further address the preferential energetics at finite temperature. To investigate the formation near the melting point, the short-range order (SRO) parameter is of great significance~\cite{Bragg}~\cite{Cowley}. However, comprehensive theoretical approach to address SRO in disordered phase of multicomponent alloy including a large number of atoms like Mg-based LPSO structure has not yet conducted.

Recently, our group found a single special microscopic state, whose structure can be constructed by information about underlying lattice, and no information about total energy is needed ~\cite{Yuge1}~\cite{Yuge2}.
This single microscopic state can characterize properties of equilibrium macroscopic states, including SRO.  
In this study, we investigate the preferential energetics of Mg-based ternary alloys in terms of SRO parameter in disordered phases of Mg-RE-Zn (RE = Y, La, Er, Ho, Dy, Tb) and Mg-Gd-Al alloys through DFT calculation using the above special state to address relationship between SRO and LPSO structure.

\section{2. Methodology}

We evaluate short-range order in disordered phases of Mg-based alloys
by calculating the statistical average of correlation function using the information
of a lattice. As our previous study~\cite{Tanaka}, obtained physical quantities by using SQSs are the arithmetic average over possible microscopic state. Furthermore, the value of the cluster function is determined regardless of the temperature and constituent elements. 
Therefore, we can ``a priori" determine a structure corresponding to SQS. 
However, note that an SQS does not contain any information of the temperature dependence of preferential energetics
that are essential to investigate the formation considered phases near the melting point, $i.e.$, we cannot evaluate SRO based on only SQS.
 
To overcome this problem, we calculated statistical average of correlation functions and the pair clusters' probability included in structures from the energies of two special microscopic states; one is SQS, and the other is a ``projection state" including the information of density of states of microscopic states in a crystalline system that is independent of temperature and interactions between constituent elements ~\cite{Yuge1}~\cite{Yuge2}.

In this approach, statistical average of correlation function, $\langle \xi \rangle_{Z}$, is described by following equation:

\begin{equation}
 \langle \xi \rangle_{Z} = \langle \xi \rangle_{1} - \sqrt{\displaystyle\frac{\pi}{2}\langle \xi \rangle_{2}}\cdot\displaystyle\frac{E_{\rm{proc.}}-E_{\rm{SQS}}}{k_{\rm{B}}T},
\end{equation}

where $\langle \xi \rangle_{1}$ and $\langle \xi \rangle_{2}$ are the moment of first and second order, $i.e.$ the average and variance of correlation function, respectively.  $E_{\rm{proc.}}$ and $E_{\rm{SQS}}$ are total energy of projection state and SQS. 

The structures of projection state and SQS are represented by correlation functions.
Let us consider a system with $N$ lattice points and $R$ components. We can represent structures by using the following correlation function,$\Psi_{\alpha}^i$:

\begin{equation}
 \Psi_{\alpha}^M = \rho_{d_1}(\sigma_i)\rho_{d_2}(\sigma_j) \cdots\rho_{d_n}(\sigma_k).
\end{equation}

Here, $\sigma_i$ is a variable that specifies the occupation of lattice point  $i$.  $\rho_{d_n}(\sigma_i)$ is the complete orthonormal basis function at lattice point $i$ , 
which is obtained by applying the Gram-Schmidt technique to the linearly independent 
polynomial set  $(1, \sigma_i, \sigma_i^2,\cdots,\sigma_i^{R-1} )$.   
$\alpha$ denotes a cluster included in the structure, $d_n$  is the index of the basis function,$\rho_(\sigma_i)$ , and  $M$ is the set of indices, $d_1d_2\cdots d_n$. In the case of a ternary alloy system, 
the occupation of a lattice point by the elements, Mg, RE (Gd), and Zn (Al) is indicated by 
$\sigma = +1, -1, 0$ respectively, leading to the basis function~\cite{basis1}:

\begin{equation}
	\left(
		\begin{array}{c}
			\rho_0(\sigma_i)\\
			\rho_1(\sigma_i)\\
			\rho_2(\sigma_i)\\\end{array}
	\right)
	=
	\left(
		\begin{array}{ccc}
			1 & 0 & 0\\
			0 & \sqrt{\displaystyle\frac{3}{2}} & 0\\
			-\sqrt{2} & 0 &\displaystyle\frac{3}{\sqrt{2}}\\
		\end{array}
	\right)
	\left(
		\begin{array}{c}
			1\\
			\sigma_i\\
			\sigma_i^2\\
			\end{array}
	\right).
\end{equation}

Combined with the constraint condition required from composition conservation law, the relation between correlation function and pair cluster's probability consisting of $I$ and $J$ elements, $p_{I-J}$, can be represented by a following matrix~\cite{Wolverton}:

\begin{equation}
	\left(
		\begin{array}{c}
			\Psi_\alpha^{11}\\[8pt]
			\Psi_\alpha^{12}\\[8pt]
			\Psi_\alpha^{22}\\[8pt]
		\end{array}
	\right)
	=
	\left(
		\begin{array}{cccccc}
			\displaystyle\frac{3}{2} & -3 & 0 & \displaystyle\frac{3}{2} & 0 & 0\\[10pt]
			\displaystyle\frac{\sqrt{3}}{4} & 0 &-\displaystyle\frac{\sqrt{3}}{2}& -\displaystyle\frac{\sqrt{3}}{4} &-\displaystyle\frac{\sqrt{3}}{2} &0\\[10pt]
			\displaystyle\frac{1}{2}& 1 & -2 & \displaystyle\frac{1}{2} & -2 & 2\\[10pt]
		\end{array}
	\right)
	\left(
		\begin{array}{c}
			p_{\rm{Mg-Mg}}\\
			p_{\rm{Mg-Y}}\\
			p_{\rm{Mg-Zn}}\\
			p_{\rm{Y-Y}}\\
			p_{\rm{Y-Zn}}\\
			p_{\rm{Zn-Zn}}\\
			\end{array}
	\right).
\end{equation}



By using this matrix, we transformed correlation functions into the pair cluster's probability.

\begin{figure}[]
	\begin{center}
		\includegraphics[width=9cm,clip]{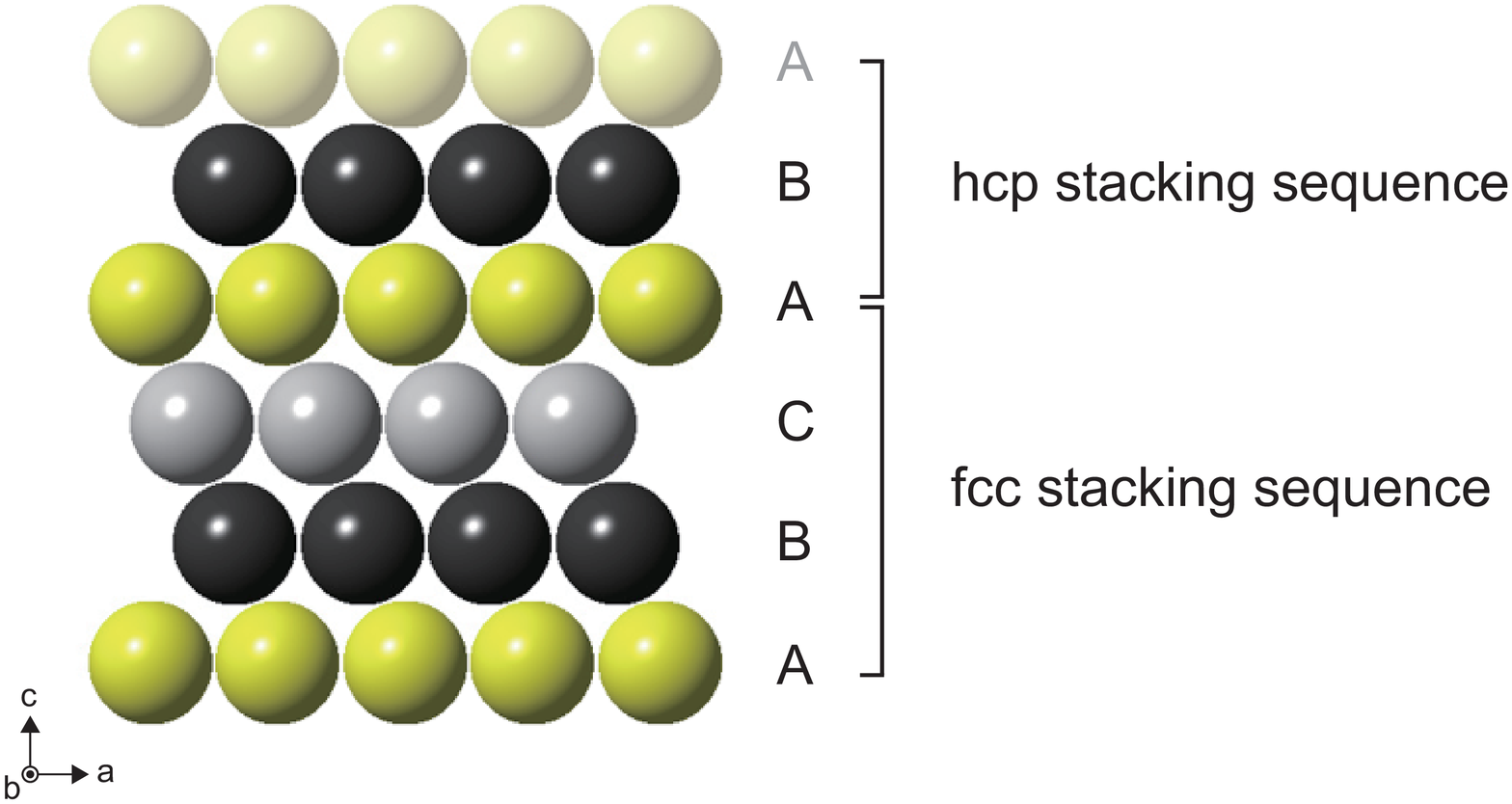}\\
		\caption{Schematic illustration of a mixed structure,
			     which is introduced stacking fault into hcp 
			     stacking sequence. A, B and C are indices 
			     of stacking sequence for fcc structure.}
		\label{fig:fig1}
	\end{center}
\end{figure}

\begin{figure}[]
	\begin{center}
		\includegraphics[width=5cm,clip]{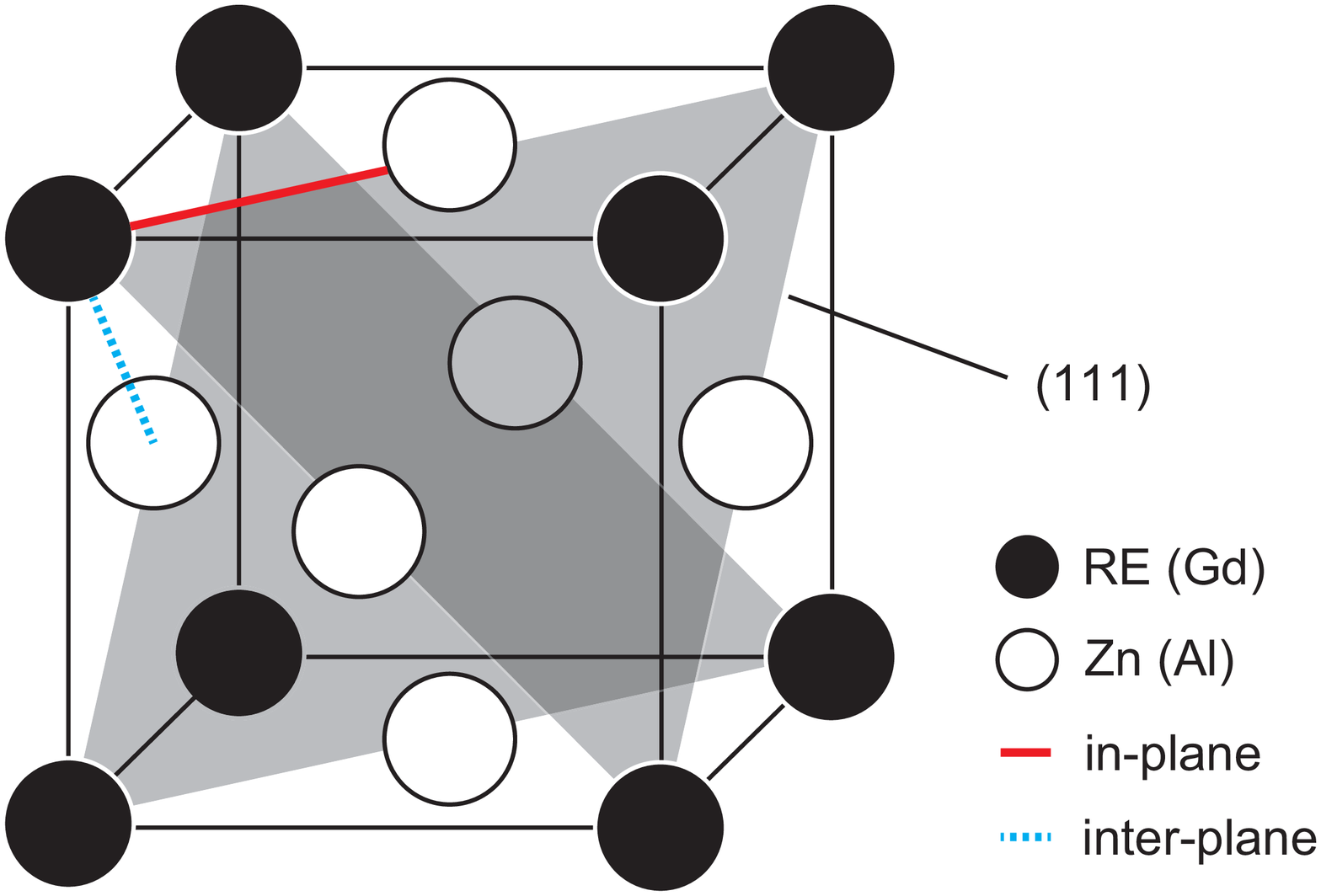}\\
		\caption{Schematic illustration of L1$_2$ cluster consisted of
		RE(Gd) and Zn(Al). Black and White spheres represent
		RE (Gd) and Zn (Al) respectively. 
		Translucence gray triangles are (111) planes for fcc structure. 
		Red and blue lines represent in-plane and inter-plane cluster's 
		bondings, respectively.}
		\label{fig:fig2}
	\end{center}
\end{figure}

As the feature of LPSO structure, we can easily point out forming long-period stacking sequence
 by introducing stacking faults into hcp stacking sequence. In this study, to investigate the effect
 of stacking fault on SRO, we calculated total energy of not only hcp structure but also a structure
 introducing stacking fault into hcp stacking sequence (ABCAB) as shown in FIG.\ref{fig:fig1},
 which we call ``mixed structure" hereinafter. A,B and C are indices for fcc stacking sequence. 
 We focus on the SRO of 1st nearest neighbor in-plane and inter-plane pair clusters, RE-Zn (Gd-Al), Zn-Zn(Al-Al), and
 Mg-Mg clusters, which construct RE-Zn (Gd-Al) L1$_2$ clusters as illustrated in FIG.\ref{fig:fig2}. 
 These clusters' probabilities were measured from those at $T$ = 1500 K.



To investigate the stability of solute pair clusters consisted of RE and Zn (Gd and Al), 
we also calculated solute energy of each pair cluster into mixed structure. 
We arranged RE-Zn (Gd-Al) pair cluster into hcp, fcc and intermediate region of pure-Mg
mixed structure and compared total energies with that of pure Mg on mixed structure.

We performed first-principles calculations using a DFT code, the Vienna Ab-initio Simulation Package (VASP)~\cite{vasp1}~\cite{vasp2}, 
to obtain the total energies for Mg-RE-Zn and Mg-Gd-Al alloys. All-electron Kohn-Sham equations were solved by employing the projector augmented-wave (PAW) method ~\cite{paw1}~\cite{paw2}. 
We selected the generalized-gradient approximation of the Perdew-Burke-Ernzerhof (GGA-PBE) form ~\cite{ggapbe}as the exchange-correlation functional. 
The plane-wave cutoff energy was set at 350 eV throughout the calculations. Brillouin zone sampling was performed on the basis of the Monkhorst-Pack scheme~\cite{monk}. 
The K-point mesh is set to 4$\times$4$\times$4 and the smearing parameter was 0.15 eV ~\cite{meth}.

\section{3. Results and Discussion}
\begin{figure*}[t]
 \begin{center}
 \includegraphics[width=18cm,bb=0 0 968 524]{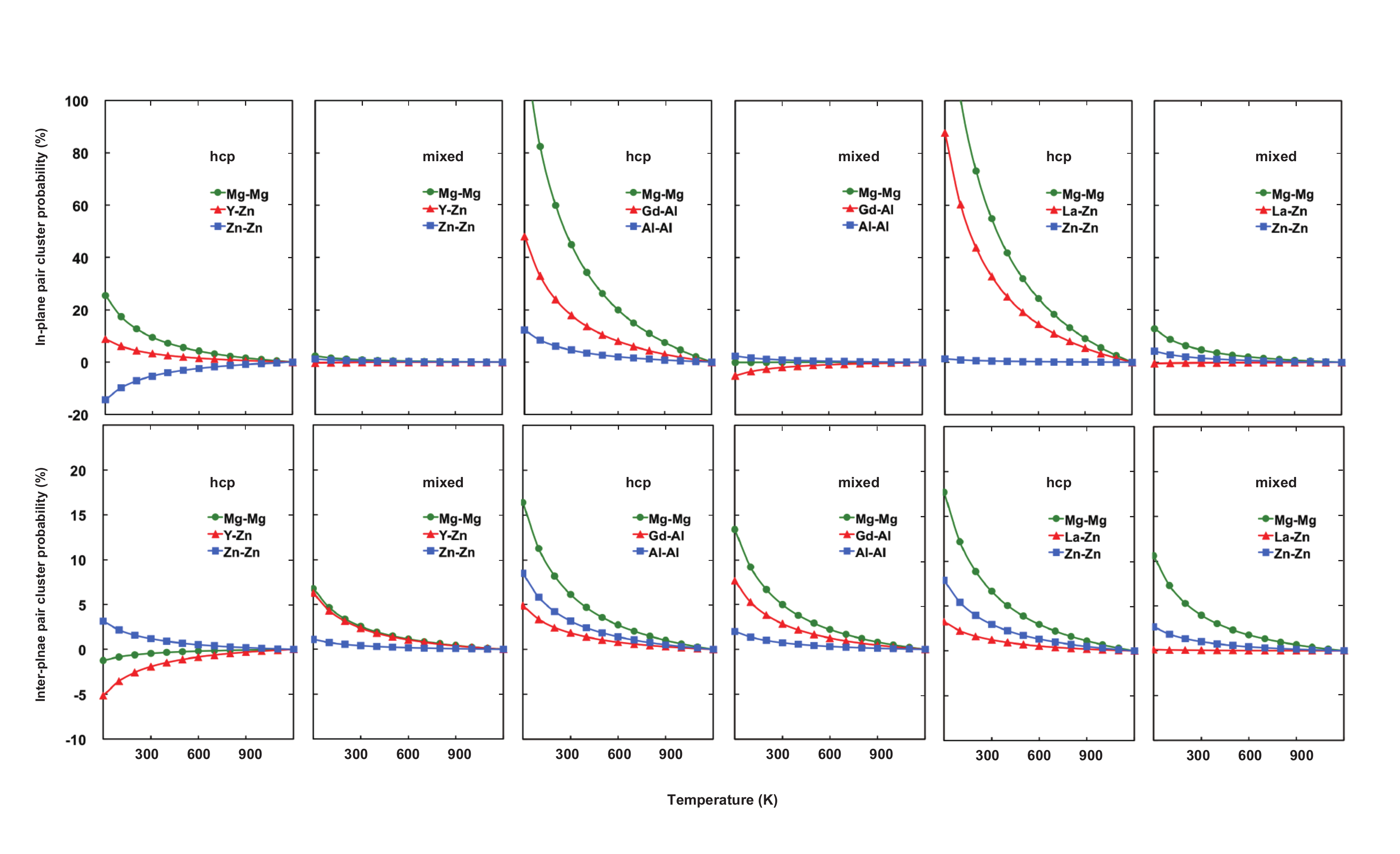}\\
		\caption{Temperature dependence of SRO in Mg-RE-Zn (RE = Y, La) and Mg-Gd-Al system on (a) hcp and (b) mixed structure.}
                 \label{fig:fig3}
 \end{center}
\end{figure*}

Figure \ref{fig:fig3} is the temperature dependence of pair clusters' probabilities of Mg-Y-Zn, Mg-Gd-Al and Mg-La-Zn system. 
First, let us focus on the inter-plane pair clusters' probability. In Mg-Y-Zn and Mg-Gd-Al system, Y-Zn and Gd-Al inter-plane pair cluster's probabilities are increasing. This is consistent with the forming L1$_2$ cluster consisted of Y(Gd) and Zn(Al). 
Moreover, the probability of Mg-Mg and Y-Zn(Gd-Al) clusters become numerically superior to that of Zn-Zn cluster after introducing stacking fault into hcp structure, which agrees well with the tendency of phase separation into Mg- and RE-Zn-rich phases along perpendicular direction. As a results of other Mg-RE-Zn system (RE = Dy, Ho, Er, Tb), RE-Zn probabilities have the same tendencies, which are also shown in Appendix. On the other hand, in Mg-La-Zn system that dose not form LPSO structure, La-Zn inter-plane pair clusters' probability of mixed structure is not increasing compared with that of disordered phases at $T$ = 1500K. (FIG.\ref{fig:fig3}). Forming RE-Zn (Gd-Al) pair cluster is a  necessary requirements for constructing L1$_2$ clusters; this preference of inter-plane pair clusters in disordered phases therefore have close relationship with the formation of LPSO structure. Then, Mg-RE-Zn LPSO system is classified by the formation process. Type 1 is  that the LPSO phase forms during solidification (RE = Y, Dy, Ho, Er in this study)\cite{Itoi}\cite{Yoshimoto} and Type 2 is that a phase precipitates from $\alpha$-Mg solid solution with soaking at 773 K (RE = Tb in this study)~\cite{Yamasaki}. These results of SRO in Mg-RE-Zn systems mean that there is a slight difference between the preference of SRO with respect to formation process of LPSO structure. 

Next, we discuss the relationship between the in-plane pair cluster's probability and formation of LPSO structure.
As shown in upper part of FIG.\ref{fig:fig3}, Y-Zn and Gd-Al in-plane pair cluster's probabilities of the mixed structure are decreasing compared with those of hcp structure, 
which is the same tendency for the other systems that form LPSO structure (FIG.\ref{fig:fig4}).
Let us consider this tendency of SRO from the viewpoint of the solute energy of pair cluster into mixed structure consisted of pure-Mg. 
Solute energies of in-plane cluster of RE(Gd) and Zn(Al) into fcc region were lower than those into hcp region. 
As a result of  this estimation of solute energy, RE-Zn (Gd-Al) clusters can be more stable on in-plane layer in fcc region than hcp region.
Thus, although the probabilities of RE-Zn (Gd-Al) in-plane clusters on a structure with only hcp stacking sequence are increasing, this tendency is considered to become opposite on the mixed structure, $i.e$, the pair cluster probabilities in hcp region will be decreasing after introducing stacking fault. On the other hand, in fcc region on the mixed structure, RE-Zn (Gd-Al) in-plane clusters are stable in terms of solute energy. The SRO in fcc region on the mixed structure should therefore be increasing, which is consistent with the forming L1$_2$ clusters in fcc region. This opposing tendency of SRO between fcc and hcp region should result in the tendency of SRO of RE-Zn (Gd-Al) in-plane cluster on the mixed structure.
In the previous study, the interfacial energy of a disordered phase with an introduced stacking fault is lower than the linear average energy for hcp and fcc stacking sequences and the interface gains ``negative" energy"~\cite{Tanaka}. Thus, introducing stacking faults into hcp region has an effect not only to the stabilizing but also the preference of SRO, which exhibits introducing stacking faults should play an essential role for the forming of LPSO structure.



\section{4. Conclusion}
To assess the formation of Mg-based LPSO structure, we systematically investigate the preference of SRO of Mg-RE-Zn (RE = Y, La, Er, Ho, Dy, Tb) and Mg-Gd-Al ternary alloy systems through first-principles calculation combined with the information of disordered state on a lattice under spatial constraint. In the system forming LPSO structure, RE-Zn and Gd-Al pair cluster probabilities are increasing due to introducing stacking faults into hcp region. In the contrast, Mg-La-Zn system that is not forming LPSO dose not show that tendency, which means that the preference of SRO of pair cluster has the profound relationship with forming the characteristic structure in LPSO.

\section{Acknowledgement}
\begin{acknowledgments}
This work is supported by a Grant-in-Aid for Scientific Research on Innovative Areas (26109710) from the Ministry of Education, Science, Sports and Culture of Japan.
\end{acknowledgments}

\appendix*
\setcounter{figure}{0}
\setcounter{table}{0}
\renewcommand{\thefigure}{A\arabic{figure}}

\section{Appendix}
\subsection{The SRO of Mg-RE-Zn (RE = Dy, Er, Ho, Tb) systems}
\begin{figure*}[t]
\rotatebox{90}{\begin{minipage}{\textheight}
	\begin{center}
		\includegraphics[width=24cm,bb=0 0 1278 526]{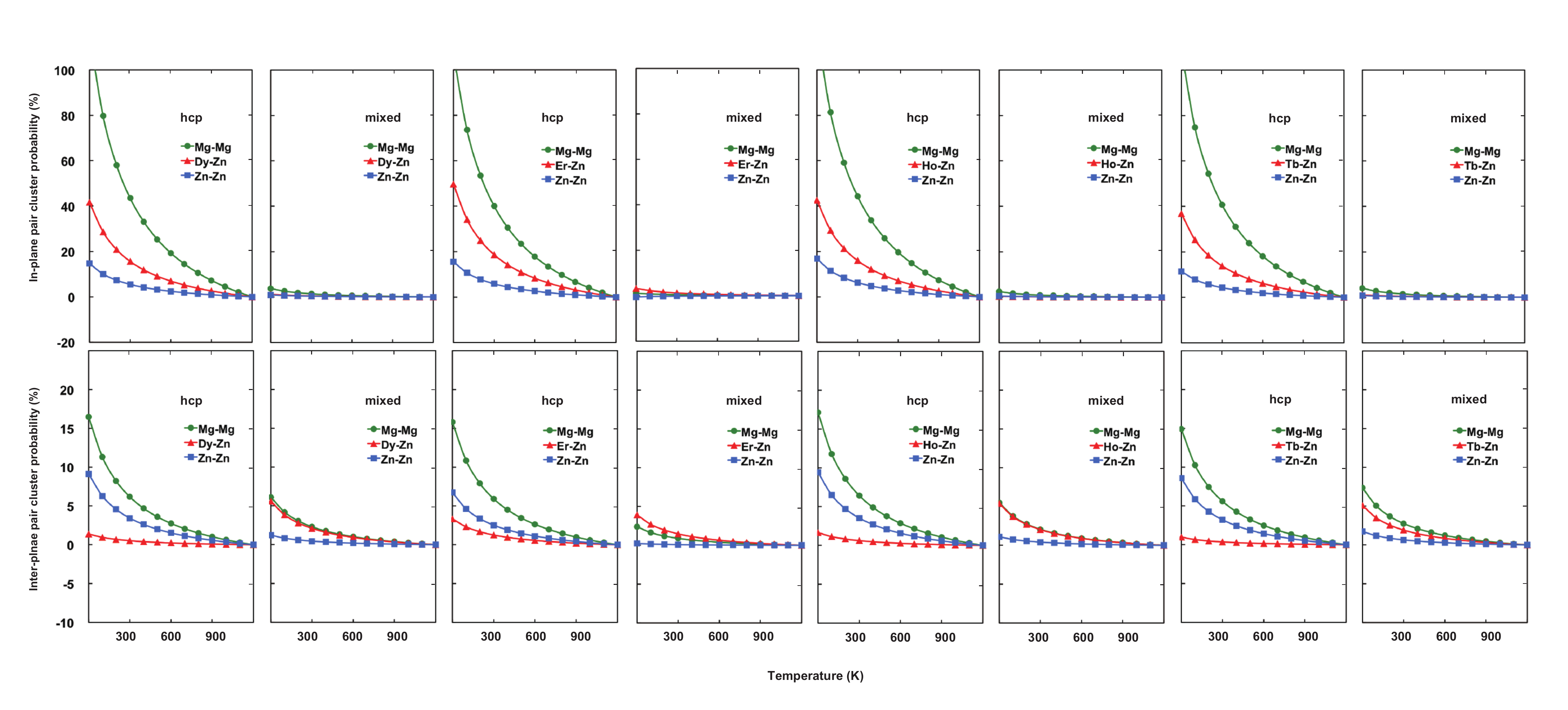}\\
		\caption{Temperature dependence of SRO in Mg-RE-Zn (RE = Dy, Er, Ho,Tb) system on (a) hcp and (b) mixed structure.}
                 \label{fig:fig4}
	\end{center}
	\end{minipage}}
\end{figure*}

The SRO of Mg-RE-Zn (RE = Dy, Er, Ho, Tb) systems are shown in FIG.\ref{fig:fig4}

\end{document}